\documentclass[twocolumn,pra,showpacs,amsmath,amssymb]{revtex4}

\usepackage{amssymb,amsmath,amsfonts,amsthm,graphicx,color}

\begin{document}

\title{Quantum Bounds on Bell inequalities}
\author{K\'aroly F.~P\'al}
\email{kfpal@atomki.hu}
\author{Tam\'as V\'ertesi}
\email{tvertesi@dtp.atomki.hu}
\affiliation{Institute of Nuclear Research of the Hungarian Academy of Sciences\\
H-4001 Debrecen, P.O.~Box 51, Hungary}

\def\CC{\mathbb{C}}
\def\RR{\mathbb{R}}
\def\one{\leavevmode\hbox{\small1\normalsize\kern-.33em1}}
\newcommand*{\tr}{\mathsf{Tr}}
\newcommand{\diag}{\mathop{\mathrm{diag}}}
\newtheorem{theorem}{Theorem}[section]
\newtheorem{lemma}[theorem]{Lemma}

\date{\today}

\begin{abstract}

We have determined the maximum quantum violation of 241 tight bipartite Bell inequalities with up to
five two-outcome measurement settings per party by constructing the appropriate measurement operators
in up to six-dimensional complex and eight-dimensional real component Hilbert spaces using numerical
optimization. Out of these inequalities 129 has been introduced here. In 43 cases higher dimensional
component spaces gave larger violation than qubits, and in 3 occasions the maximum was achieved with
six-dimensional spaces. We have also calculated upper bounds on these Bell inequalities using a method
proposed recently. For all but 20 inequalities the best solution found matched the upper bound.
Surprisingly, the simplest inequality of the set examined, with only three measurement settings per
party, was not among them, despite the high dimensionality of the Hilbert space considered. We also
computed detection threshold efficiencies for the maximally entangled qubit pair. These could be
lowered in several instances if degenerate measurements were also allowed.

\end{abstract}

 \maketitle

\section{Introduction}\label{intro}

Suppose two classical systems which are separated from each other. Let us make some local measurements
on them. Then the results of these measurements may be correlated, which may be explained by shared
randomness experienced in the past. For a given number of possible measurement choices (inputs) and
results (outputs) the set of correlations forms a polytope whose finite number of vertices correspond
to all the deterministic assignments of outputs to inputs. The facets of the polytope, which form the
boundary of the classical region, correspond to tight Bell inequalities \cite{Bell64,CHSH}. Thus Bell
inequalities has no a priori relation to quantum physics. However, quantum physics violates them,
which has been verified experimentally in numerous occasions up to some technical loopholes (see e.g.,
\cite{exp}). Indeed, this makes Bell inequalities very interesting.

On the other hand, one may ask what is the achievable set of correlations if one allows the two
parties to share quantum resources as well over shared randomness. This is a convex set, such as for
the classical case, however it cannot be described by a finite number of extreme points \cite{KT92}.
Nevertheless, one can construct so called quantum-Bell inequalities which bound the correlations
achievable by quantum physics \cite{KT92, Tsirel93, Tsirel06} (see also item~5 in Sec.~III.C of
Ref.~\cite{AS12}). A simple way to form quantum-Bell inequalities is to use the coefficients of known
Bell inequalities, and determine the maximum value one can get by performing measurements on quantum
systems of arbitrary dimensions.

The simplest result in this respect is the Tsirelson bound \cite{Tsirel80} stating that quantum
correlations cannot violate the CHSH inequality \cite{CHSH} by more than $(\sqrt 2 -1)/2$. Note,
however that the requirement of non-signaling alone allows a higher bound of 1 \cite{KT85,PR94} (the
maximum allowed value in a local classical theory is 0). In the present paper we stick to the
bipartite two-outcome scenario and we consider inequalities with more than two measurement settings
having nontrivial local marginals as well.

In particular, in this paper we test 241 bipartite tight Bell
inequalities with two-valued observables with up to five settings.
These Bell inequalities have been collected from various sources:
from the list of Ref.~\cite{89list} containing 89 inequalities
labelled by $A_i$, $i=3,\ldots,89$, from the list of
Ref.~\cite{BrunGis} comprising 31 inequalities with up to four
settings per party. 18 of them, labelled by $I_{4422}^i$,
$i=3,\ldots,20$ are introduced by that paper. The simplest
nontrivial inequality $I_{3322}$ beside the CHSH inequality was
presented first by Froissart \cite{froissart} in 1981 and more
recently it was reintroduced by Sliwa \cite{Sliwa} and by Collins
and Gisin \cite{14422}. The rest of them, namely,
$I_{3422}^{1,2,3}$ and $I_{4422}^{1}$ were presented in
Ref.~\cite{14422}, $I_{4422}^{2}$ in \cite{24422}, $A_{5,6}$ and
$AII_{1,2}$ in \cite{A5A6AII12}, while $AS_{1,2}$ in \cite{AS12}.
These lists were also considered by us previously \cite{PV08}.
Finally, a list of 129 tight bipartite Bell inequalities with four
settings per party is constructed in the present paper (labelled
by $J_{4422}^i$, $i=1,\ldots,129$). We test numerically these Bell
inequalities and give the bound on them achievable by quantum
systems, thereby generating quantum-Bell inequalities. On one
hand, we used numerical optimization in higher dimensional (up to
six complex and eight real) component Hilbert spaces to calculate
a lower bound on maximal Bell violation. On the other hand, the
hierarchy of tests of Refs.~\cite{NPA07,NPA08} allows us to give
upper bounds on the violation achievable by quantum systems. We
found in a surprisingly many cases that the lower and upper bounds
coincide indicating that the quantum bound we have found is tight.
Numerics suggests that in some cases even five-dimensional local
Hilbert spaces (i.e., 25-dimensional full Hilbert space) were not
enough to achieve the maximum quantum value. This result can be
interpreted through the recent concept of dimension witness
\cite{Brunner}, which is a certain kind of generalization of Bell
inequalities which enables to measure the dimension of the Hilbert
space (see also Refs.~\cite{Perez,WCD} for recent works using this
concept). In particular, a $d$-dimensional Hilbert space witness
allows to distinguish the strength of correlations of measurement
outcomes on members of composite systems that can be obtained in
$d$-dimensional component Hilbert spaces from the ones which are
achievable only in higher dimensions. Our result thus indicates a
five-dimensional witness, which improves on our recent result
\cite{PV08}, where we found three-dimensional witness. We also
made a study, given these tight Bell inequalities, from a
different perspective. Namely, we computed the threshold detection
efficiency for a maximally entangled qubit pair required to close
the detection loophole in the symmetric (Alice and Bob having the
same detection efficiency) and in the asymmetric cases (Alice has
perfect detection efficiency) by considering degenerate
measurements (having deterministic outcomes) as well.
Surprisingly, in almost all cases degenerate measurements were
required to achieve the lowest threshold efficiency in both the
symmetric and asymmetric situations. However, let us note, that
these cases with degenerate measurements could be traced back to
ones with non-degenerate measurements, which achieve the same
threshold value. In the asymmetric case we could improve on the
best known results from the literature \cite{BrunGis,BGSS}, by
carrying out only non-degenerate measurements.

The rest of the paper is organized as follows. In
Sec.~\ref{method} we describe methods to find tight two-outcome
bipartite Bell inequalities, and we show the 129 inequivalent
examples we have got this way. For all of them the classical limit
is zero, and we have given a symmetric form, whenever there exists
one. In Sec.~\ref{lower} it is explained how we constructed the
maximum quantum violating solution in real and complex six-, and
real eight-dimensional component Hilbert spaces. In
Sec.~\ref{upper} we present briefly the method of Navascu\'{e}s et
al.\ \cite{NPA07,NPA08} to obtain upper limit for the maximum
quantum violation of two-party two-outcome Bell inequalities. In
Sec.~\ref{disc} we give the maximum quantum value we have derived,
the size of the Hilbert spaces required, and the dimensionalities
of the projectors used as the measurement operators for all the
241 Bell inequalities considered. We also show when this value
reaches the upper bound. For those cases the value given is the
absolute quantum bound. We discuss the results obtained. In
Section~\ref{detect} the minimal detection efficiencies are
computed for a maximally entangled pair of qubits considering
degenerate measurements as well. More details of the results
including the matrices of the optimal operators and the Schmidt
coefficients of the states are presented in the web
site~\cite{Bellweb}. Section~\ref{conc} summarizes the results of
the paper.

\section{Methods for generating tight Bell inequalities}\label{method}

We consider a typical experiment to test for correlations. Imagine a system, either quantum or
classical, which is composed of two subsystems, $A$ and $B$. Suppose that Alice has $m_A$ choices of
two-valued measurements on $A$, while Bob has $m_B$ choices of two-valued measurements on $B$. Let us
refer to this situation with the notation $m_Am_B22$. The outcome of the measurements is one of two
values labelled by the symbols $1$ and $0$. The experiment is repeated many times with different
choices of measurement settings so as to get accurate probabilities. The result of such a correlation
experiment can be thought of as a vector $\vec p$ in a $d$ dimensional space, where $d=m_A+m_B+m_A
m_B$. The components $p_{A_i}$, $p_{B_j}$ and $p_{A_i B_j}$ (for $1\le i\le m_A$ and $1\le j\le m_B$)
of $\vec p$ represent the probability that Alice gets 1 for measurement setting $i$, Bob gets 1 for
measurement setting $j$, and that both Alice and Bob get 1 for setting $i$ and $j$ on the respective
sides.

In a quantum correlation experiment Alice and Bob share a quantum state $\rho$, and the probabilities
can be expressed by the formulae
\begin{align}
p_{A_i}&=\tr\left[\rho\left(A_i\otimes I_B\right)\right],\nonumber\\
p_{B_j}&=\tr\left[\rho\left(I_B\otimes B_j\right)\right],\nonumber\\
p_{A_i B_j}&=\tr\left[\rho\left(A_i\otimes B_j\right)\right], \label{prob}
\end{align}
where $I_A$ $(I_B)$ denotes the unity operator acting on the Hilbert space of Alice (Bob), while $A_i$
and $B_j$ are the $\{1,0\}$ valued observables (i.e., projectors) measured by Alice and Bob,
respectively. Note, that for our case of binary outcomes it has been shown \cite{CHTW04} that
projective measurements can reproduce the set of all quantum correlations.

On the other hand, the result of a classical correlation experiment must correspond to a probability
distribution represented as a convex sum over all deterministic configurations. A deterministic
configuration assigns the outcomes $\{1,0\}$ to each of the $m_A + m_B$ measurements, in which case
the joint probabilities factor into two local probability distributions, $p_{A_i B_j}=p_{A_i}p_{B_j}$.
Thus, the set of vectors $\vec p$ which are possible results of a correlation experiment forms a
$d=(m_A+m_B+m_A m_B)$-dimensional polytope $P$ \cite{Pitowsky}. Froissart \cite{froissart} has shown
that in the general $m_A m_B22$ case the set of all $n=2^{m_A+m_B}$ points $\vec p_v$, $v=1,\ldots,n$
corresponding to deterministic configurations are actually the extremal vertices of this
$d$-dimensional polytope $P$. The experimental result has a locally causal model if and only if the
corresponding point is located inside the polytope $P$. A Bell inequality is a linear inequality
satisfied by all the points in such a polytope. A tight Bell inequality, which is the most useful to
detect non-local correlations, defines a facet of this polytope. In our particular formulation we
represent a Bell inequality in the form
\begin{equation}
\sum_{i=1}^{m_A}{b_{A_i}p_{A_i} +
\sum_{j=1}^{m_B}{b_{B_j}p_{B_j}}+\sum_{i=1}^{m_A}\sum_{j=1}^{m_B}{b_{A_i B_j}p_{A_i B_j}} \le b_0},
\label{Bell}
\end{equation}
where the sums run over all measurements of Alice and Bob, and the coefficients $b$'s are suitably
chosen real numbers. The simplest nontrivial tight Bell inequality (with the smallest values of
$m_A=2$ and $m_B=2$) is the CHSH inequality \cite{CHSH}
\begin{equation}
-p_{A_1}-p_{B_1}+p_{A_1B_1}+p_{A_1B_2}+p_{A_2B_1}-p_{A_2B_2}\le 0.
\end{equation}
Also, note that for a given Bell inequality $\vec{b} \vec{p} \leq b_{0}$, with $\vec p \in \RR^d$ and
for a $d$-dimensional polytope $P$, the face represented by the inequality is a facet of that polytope
$P$ if and only if the rank of the matrix containing the vertices $p_v$ of polytope $P$ which saturate
the inequality is $d-1$. This fact enables us to test tightness of Bell inequalities.

Given the vertices $p_v$ of polytope $P$ in case $m_A m_B 22$, our task is to find some facets, which
defines tight Bell inequalities. A deterministic algorithm (see for instance the package cdd
\cite{cdd}), which explores systematically all the facets, runs too slowly already for the 4422 case
to complete this task. This fact was also mentioned by Collins and Gisin in Ref.~\cite{14422}. In
fact, it has been shown by Pitowsky that this problem for the general $m_A m_B 22$ case is NP-complete
\cite{Pit89}. Therefore, we turn to heuristic algorithms and provide two methods which enable us to
produce tight two-outcome bipartite Bell inequalities with $m_A=m_B=4$ measurement settings. The first
one uses some portions from Seidel's shelling algorithm \cite{Seidel}, while the second one is based
on a reduction of the polytope $P$ to smaller ones. Actually, the second method proves to be useful
for a larger number of measurement settings as well. For instance, we found that it also works for the
5522 case.

\subsection{Algorithm: shelling of a Bell polytope}\label{shelling}

In this subsection we suitably modify a part of Seidel's shelling algorithm \cite{Seidel}. Consider a
$d$-dimensional polytope $P$ corresponding to the case $m_A m_B22$ (i.e., a Bell polytope), and let
$S$ denote the set of vectors $\vec p_v$ corresponding to deterministic configurations as defined
earlier. Imagine travelling along a directed straight line $L$ that intersects the interior of $P$.
Start at a point on $L$ that is in the interior of $P$ and move along $L$. Continue moving until you
pass through some facet $F$ and then leave the interior of $P$. This facet $F$ is the first in the
shelling order. Continue moving away from $P$ along $L$ and more and more facets will become visible.
The order in which these facets appear is the shelling order. This algorithm is deterministic in
essence, which makes a shelling of the polytope by travelling along a line and determine the facets of
the convex hull as they become visible. However, this algorithm runs too long to complete the task in
a reasonable time for polytope $P$ of the case 4422. Therefore, a heuristic algorithm is given which
uses a part of Seidel's algorithm:

Suppose we know a priori a single facet $F$ of the polytope $P$ defined by all vertices $\vec p_v^F
\in S_F$ which lie on the supporting hyperplane of this facet $F$. Define $\vec a$ by adding up these
vectors $\vec p_v^F$. This sets the direction of the shelling line $L$, parameterized by $x(t)=-\vec
a/t$, $-\infty<t<+\infty$. Linear programming can be used to determine for every $\vec p_v\in S$ the
first facet in the shelling that contains $\vec p_v$. Thus for each $\vec p_v\in S$ one obtains a
tight Bell inequality defined by a facet of $P$. Let us collect these facets, which serve as new
inputs to the algorithm. This routine is iterated as long as no more new Bell inequalities are found.
However, several of them may turn out to be equivalent, i.e., symmetric under party exchange,
observable exchange or relabelling of outcomes (see e.g., \cite{Mas02}). Thus one has to take care of
selecting the inequivalent Bell inequalities.

 \begin{table*}[t]
 \caption{Tight bipartite Bell inequalities with four two-outcome measurement settings
 per party, generated with the methods of Section~\ref{method}. The coefficients,
 positive or negative single-digit integers, are
 shown in the order of $b_{A_1}$, $b_{A_2}$, $b_{A_3}$, $b_{A_4}$, $b_{B_1}$,
 $b_{B_2}$, $b_{B_3}$, $b_{B_4}$, ${b_{A_1 B_1}}$, ${b_{A_1 B_2}}$,
 ${b_{A_1 B_3}}$, ${b_{A_1 B_4}}$, ${b_{A_2 B_1}}$, ${b_{A_2 B_2}}$,
 ${b_{A_2 B_3}}$, ${b_{A_2 B_4}}$, ${b_{A_3 B_1}}$, ${b_{A_3 B_2}}$,
 ${b_{A_3 B_3}}$, ${b_{A_3 B_4}}$, ${b_{A_4 B_1}}$, ${b_{A_4 B_2}}$,
 ${b_{A_4 B_3}}$, ${b_{A_4 B_4}}$ (see Eq.~(\ref{Bell})). For each inequality
 only one of the equivalent forms are shown. The form chosen has classical value zero,
 and whenever possible, it is symmetric.}
 \vskip 0.2truecm
 \centering
 \begin{tabular}{l l l l l l}
 $J^{1}_{4422}$&0-200-2-2-201-11-12121-111011-10&\hphantom{}$J^{44}_{4422}$&-2-1000-2-101111101-1010-1-2110&\hphantom{}$J^{87}_{4422}$&00-1-30-3-3-1-2211011-21-2212311\\
 $J^{2}_{4422}$&00-1-3-2-20-3-210221-211-2112221&\hphantom{}$J^{45}_{4422}$&-3-200-5-2002212212-212-1-14-3-10&\hphantom{}$J^{88}_{4422}$&00-1-2-1-2-30-1110111-21-2211211\\
 $J^{3}_{4422}$&-2-30-5-10-1-3-321331-321-2113221&\hphantom{}$J^{46}_{4422}$&-2-10-5-20-1-4-221211-321-4133322&\hphantom{}$J^{89}_{4422}$&00-2-40-3-4-2-3212-112-22-3323411\\
 $J^{4}_{4422}$&-1-20-4-10-1-3-211221-321-2112221&\hphantom{}$J^{47}_{4422}$&-2-10-4-20-1-3-221210-221-3123311&\hphantom{}$J^{90}_{4422}$&-10-1-3-10-1-32-311-3-21211-121221\\
 $J^{5}_{4422}$&-3-40-3-10-1-2-231222-221-3122210&\hphantom{}$J^{48}_{4422}$&0-6-20-2-20-1-11-112134122-321-20&\hphantom{}$J^{91}_{4422}$&-5-20-1-5-20-1223323-423-4-11321-3\\
 $J^{6}_{4422}$&-2-2-200-2001101-1111121-1-12-10&\hphantom{}$J^{49}_{4422}$&-1-3-50-4-10-11-111313-4122331-31&\hphantom{}$J^{92}_{4422}$&-5-20-1-5-20-1123323-423-4-11321-2\\
 $J^{7}_{4422}$&-2-30-4-10-1-1-221122-211-2112220&\hphantom{}$J^{50}_{4422}$&-200-4-20-1-3-22121-1-221-2113311&\hphantom{}$J^{93}_{4422}$&-20-2-40-2-30-2212-122-31-2312311\\
 $J^{8}_{4422}$&0-2000-4-2-2-12-112222-121-2-2-111&\hphantom{}$J^{51}_{4422}$&0-40-1-2-3-10111-2312211-10-3211&\hphantom{}$J^{94}_{4422}$&-20-1-3-2-1-20-2211111-21-2212211\\
 $J^{9}_{4422}$&-10-1-10-2-3-21101-2122112-21-111&\hphantom{}$J^{52}_{4422}$&-2-4-200-3-4-2222-2313312-11-6321&\hphantom{}$J^{95}_{4422}$&-30-2-6-2-1-40-3321112-31-4423422\\
 $J^{10}_{4422}$&-20-1-20-6-4-323-12-3333132-32-221&\hphantom{}$J^{53}_{4422}$&-2-4-200-3-2-2221-3312312-11-4311&\hphantom{}$J^{96}_{4422}$&-20-1-5-2-1-40-3221112-31-3313322\\
 $J^{11}_{4422}$&0-2000-4-2-3-12-122222-121-2-1-111&\hphantom{}$J^{54}_{4422}$&-2-100-3-2001211101-112-1-12-200&\hphantom{}$J^{97}_{4422}$&-6-30-1-6-30-1124324-434-4-11331-2\\
 $J^{12}_{4422}$&-30-3-3-30-3-3231-23-33313-22-2321&\hphantom{}$J^{55}_{4422}$&-3-20-1-5-3001321202-223-2-14-301&\hphantom{}$J^{98}_{4422}$&-20-2-50-2-50-3222-123-31-3413321\\
 $J^{13}_{4422}$&-3-300-3-30023123-33-213-1-22-2-20&\hphantom{}$J^{56}_{4422}$&0-2-3-4-20-3-11-431112-2-23213311&\hphantom{}$J^{99}_{4422}$&-20-3-4-200-3-21121-1-332-2322311\\
 $J^{14}_{4422}$&-1-10-4-10-1-2-211111-211-2112221&\hphantom{}$J^{57}_{4422}$&-2-40-2-2-300-121122-3413-2-42120&\hphantom{}$J^{100}_{4422}$&00-2-50-4-6-2-3212-123-42-3423522\\
 $J^{15}_{4422}$&-2-20-5-10-1-2-321221-211-2113221&\hphantom{}$J^{58}_{4422}$&-2-10-2-2-10-241-311-111-31-221121&\hphantom{}$J^{101}_{4422}$&0-2-4-3-5-20032-1-222-2212314-421\\
 $J^{16}_{4422}$&-4-200-3-2002213202-212-1-12-2-10&\hphantom{}$J^{59}_{4422}$&-1-2-200-2-3-2111-1202212-21-3221&\hphantom{}$J^{102}_{4422}$&-3-100-3-1003112111-211-2-12-2-1-1\\
 $J^{17}_{4422}$&0-4-1-10-4-1-1-2211202212-11121-3&\hphantom{}$J^{60}_{4422}$&0-2-1-10-2-1-1-311110111101111-1&\hphantom{}$J^{103}_{4422}$&0-3-4-4-5-20042-1-322-2312314-422\\
 $J^{18}_{4422}$&-3-3-30-3-3-3013-223033-2321231-5&\hphantom{}$J^{61}_{4422}$&0-4-1-10-4-1-1-4211212212-11121-2&\hphantom{}$J^{104}_{4422}$&00-3-60-4-7-3-4213-224-43-4534622\\
 $J^{19}_{4422}$&-1-4-10-1-4-1012-112022-1211121-4&\hphantom{}$J^{62}_{4422}$&-1-4-10-1-4-10-3211212212-11121-3&\hphantom{}$J^{105}_{4422}$&-3-30-2-3-30-2412-2112322-42-232-1\\
 $J^{20}_{4422}$&-4-200-2-1002113202-211-1-11-1-10&\hphantom{}$J^{63}_{4422}$&00-2-2-10-1-31-201-201211-122111&\hphantom{}$J^{106}_{4422}$&00-1-40-2-4-1-2111-112-21-2213321\\
 $J^{21}_{4422}$&-2-30-4-10-1-3-221331-321-2112220&\hphantom{}$J^{64}_{4422}$&-2-400-1-4-1-2121-1323212-31-4211&\hphantom{}$J^{107}_{4422}$&00-1-40-2-4-2-3112-112-21-2213322\\
 $J^{22}_{4422}$&-4-200-4-2002213212-212-1-13-2-10&\hphantom{}$J^{65}_{4422}$&-2-4000-4-3-2121-14232-12-31-5231&\hphantom{}$J^{108}_{4422}$&-200-4-200-422-322-3-11-3-1-132133\\
 $J^{23}_{4422}$&-400-100-1-22212-10-110-211-2111&\hphantom{}$J^{66}_{4422}$&-10-3000-1-21-211-10-111212-2-111&\hphantom{}$J^{109}_{4422}$&-8-20-2-8-20-234534-62252-1-432-44\\
 $J^{24}_{4422}$&0-1-2-2-4-20032-1-210-1112113-211&\hphantom{}$J^{67}_{4422}$&-2-200-1-1-40121-121211-22-1-3120&\hphantom{}$J^{110}_{4422}$&-10-2-3-10-2-3-21121-1-321-3322221\\
 $J^{25}_{4422}$&-300-100-1-2211100-110-111-2111&\hphantom{}$J^{68}_{4422}$&-2-4000-4-2-2121-1323202-31-4221&\hphantom{}$J^{111}_{4422}$&0-1-3-1-3-10-121-1-111-2112212-211\\
 $J^{26}_{4422}$&-3-100-3-1003112101-111-1-12-1-1-1&\hphantom{}$J^{69}_{4422}$&0-2-1-2-20-101-1101011111-221-11&\hphantom{}$J^{112}_{4422}$&-2-300-3-10-2211-2122211-212-201\\
 $J^{27}_{4422}$&-1-2-30-1-2-30211-21-13213-12-222-2&\hphantom{}$J^{70}_{4422}$&-20-4000-2-32-322-10-221322-3-121&\hphantom{}$J^{113}_{4422}$&-5-2-10-5-2-10123322-323-301321-3\\
 $J^{28}_{4422}$&-4-200-4-2003213202-212-1-13-2-1-1&\hphantom{}$J^{71}_{4422}$&-500-100-1-2222200-110-211-2111&\hphantom{}$J^{114}_{4422}$&-60-3-3-1-4-104222121-332-21-5413\\
 $J^{29}_{4422}$&-5-300-5-30043143-13-213-1-24-2-2-2&\hphantom{}$J^{72}_{4422}$&-20-3000-2-32-221-10-221222-3-121&\hphantom{}$J^{115}_{4422}$&-50-2-1-2-4-303233222-422-31-3311\\
 $J^{30}_{4422}$&-2-30-3-10-2-1-121122-211-221122-1&\hphantom{}$J^{73}_{4422}$&-3-5-100-5-2-3231-2324313-42-5321&\hphantom{}$J^{116}_{4422}$&-40-1-1-1-3-202222121-311-21-2211\\
 $J^{31}_{4422}$&-10-1-30-4-2-112-11-2221121-22-121&\hphantom{}$J^{74}_{4422}$&-1-200-1-10-311111011-11-121-1-11&\hphantom{}$J^{117}_{4422}$&-5-1-2-10-5-102322131-313-31-3311\\
 $J^{32}_{4422}$&-1-2-10-1-2-100011032-212-111-21-1&\hphantom{}$J^{75}_{4422}$&-2-20-100-3-11111112-10-221-2111&\hphantom{}$J^{118}_{4422}$&-3-1-1-10-3001211021-212-20-2201\\
 $J^{33}_{4422}$&-3-10-200-1-2121101-110-312-1211&\hphantom{}$J^{76}_{4422}$&0-1-4-2-3-20022-1-211-2113213-211&\hphantom{}$J^{119}_{4422}$&-1-4-1-2-6-200321-2122232-314-411\\
 $J^{34}_{4422}$&-2-20-3-20-1-2-121111-221-212321-1&\hphantom{}$J^{77}_{4422}$&-40-2-10-4-102221021-213-21-3211&\hphantom{}$J^{120}_{4422}$&-2-50-2-6-20-1322-3123332-314-411\\
 $J^{35}_{4422}$&00-1-30-3-1-3-221101-111-3122311&\hphantom{}$J^{78}_{4422}$&-3-30-10-4-1-2132-2212203-31-3221&\hphantom{}$J^{121}_{4422}$&-20-4-20-30012-1102-1-22220-3311\\
 $J^{36}_{4422}$&-3-20-200-1-3121102-220-313-1211&\hphantom{}$J^{79}_{4422}$&-20-6000-2-32-422-10-221433-3-221&\hphantom{}$J^{122}_{4422}$&-30-2-1-2-2-302122212-321-21-2211\\
 $J^{37}_{4422}$&-2-10-1-30-2-1021120-211-221211-1&\hphantom{}$J^{80}_{4422}$&-2-3-200-2-401211312-21-231-423-1&\hphantom{}$J^{123}_{4422}$&-10-6-5-1-4-1012-21121-34222-5433\\
 $J^{38}_{4422}$&00-3-4-10-3-61-2-12-403412-234132&\hphantom{}$J^{81}_{4422}$&-10-4000-1-21-21100-111222-2-111&\hphantom{}$J^{124}_{4422}$&-200-5-30-2-2-32121-1-112-3214411\\
 $J^{39}_{4422}$&0-1-3-3-10-2-51-2-12-303412-233121&\hphantom{}$J^{82}_{4422}$&-2-2-200-3-1-1121-1211112-11-4211&\hphantom{}$J^{125}_{4422}$&-20-2-3-20-2-34-412-4-21311-122321\\
 $J^{40}_{4422}$&-2-4-300-4-3-2231-3313313-22-5321&\hphantom{}$J^{83}_{4422}$&00-1-30-2-3-2-2211001-11-3222311&\hphantom{}$J^{126}_{4422}$&-1-10-4-20-3-1-1111112-31-3212222\\
 $J^{41}_{4422}$&-1-4-10-1-4-10-32112122122-212-20&\hphantom{}$J^{84}_{4422}$&0-1-6-1-6-30-133-1-211-3224334-411&\hphantom{}$J^{127}_{4422}$&00-30-30-2-4-3-1121-1-1233222-310\\
 $J^{42}_{4422}$&-1-40-1-1-40-1-3211212212-21121-2&\hphantom{}$J^{85}_{4422}$&-4-30-1-4-30-1113214-333-3-11231-2&\hphantom{}$J^{128}_{4422}$&00-50-5-2-1-321-1-211-3234432-311\\
 $J^{43}_{4422}$&-10-300-1-2-1111-10-1111202-2110&\hphantom{}$J^{86}_{4422}$&-3-20-1-3-20-1012213-322-3-11221-1&\hphantom{}$J^{129}_{4422}$&00-2-50-2-4-1-3121012-32-2213322\\

 \end{tabular}
 \label{table:newineqs}
 \end{table*}

\subsection{Algorithm: slicing a Bell polytope}\label{cutting}

Similarly as in \ref{shelling} we start from an a priori known facet $F$ of a $d$-dimensional polytope
$P$ corresponding to a tight Bell inequality, $\vec b \vec p\le b_0$, where $p\in R^d$. In this case
for all vertices $\vec p_v \in S$ corresponding to deterministic configurations the inequality holds
by definition. Now let us decrease the value of $b_0$ to a smaller value $b_0^*$. Thereby we obtain
two groups of vertices, one ($S'$) for which $\vec b\vec p\ge b_0^*$ holds and one ($S''$) for which
it does not hold.

The convex hull of these vertices form polytopes, designated by $P'$ and $P''$, respectively. By
setting an appropriate value for $b_0^*$ we may get a polytope $P'$ in the same dimension $d$ as for
the polytope $P$. Now we take as an input these vertices to the cdd program \cite{cdd} which generate
all the facets of this reduced polytope $P'$ (assuming that the number of vertices in $P'$ is small
enough to be handled by the program). However, not all of the generated facets will be the facets of
the original polytope $P$ as well. Thus one must sort out the valid faces which fulfill $\vec b'\vec
p_v\le b_0'$ for all $\vec p_v\in S$, with $\vec b'$ and $b_0'$ defining a facet of $P'$. Such as in
the algorithm \ref{shelling}, these facets, which define tight Bell inequalities, form the input to
the new cycle of computations. The computation halts only when no more new Bell inequalities can be
produced this way.

\subsection{Comments specific to the 4422 Bell inequalities}

A crucial point of both algorithms is the right choice of the initial facet. A facet in the
$d$-dimensional polytope $P$ is called degenerate if it is created by more than $d$ points. The
problem lies in the fact that a Bell polytope $P$ in general, and also in the particular case of 4422
is highly degenerate. However, it is not favorable to start the algorithm \ref{cutting} from a highly
degenerate facet, since in this case the number of vertices of $P'$ which are fed in the cdd program
can easily become larger than the number the program can handle. In our specific case we used Brunner
and Gisin's $I_{4422}^{19}$ Bell inequality as an input facet to the algorithm and the 129 tight Bell
inequalities were obtained using both methods \ref{shelling} and \ref{cutting}. However, we mention
that we did not perform an exhaustive search, since it would have been time consuming. We suspect that
the list of 26 inequalities of Ref.~\cite{BrunGis} plus our 129 introduced ones do not form the
complete set of all 4422-type of tight Bell inequalities. This is implied by the fact, that our set of
129 inequalities covered only 23 from the list of 26 inequalities of Brunner and Gisin. The
coefficients defining the 129 inequalities are shown in Table~\ref{table:newineqs}.

\section{Lower bounds}\label{lower}

The quantum value of a Bell expression is the expectation value of
the operator
\begin{align}
&\sum_{i=1}^{m_A}{b_{A_i}\left(A_i\otimes I_B\right)}
+ \sum_{j=1}^{m_B}{b_{B_j}\left(I_B\otimes B_j\right)}\nonumber\\
&+ \sum_{i=1}^{m_A}\sum_{j=1}^{m_B}{b_{A_iB_j}\left(A_i\otimes B_j\right)}.
\label{Bellquant}
\end{align}
According to Eqs.~(\ref{prob}) and (\ref{Bell}) this is the value one
gets for the left hand side of Eq.~(\ref{Bell}) with measurements on
a quantum system. This can be larger then the maximum classical
value $b_0$, the quantum violation is the difference of the
quantum value and $b_0$. For the maximum quantum violation it is
enough to consider pure states.

We have given lower limits for the maximum violation of Bell inequalities by explicitly constructing
measurement operators in Alice's and Bob's component Hilbert spaces and the appropriate state using
numerical optimization. We have made calculations with six-dimensional real and complex spaces and
eight-dimensional real spaces, taking three- and four-dimensional projection operators as measurement
operators, respectively. The results of such a calculation may correspond to solutions in component
spaces of lower dimensionality. The six-dimensional case incorporates the five-dimensional case, in
which each measurement operator can be a two- or a three-dimensional projector, the four-dimensional
case with one-, two- and three-dimensional projectors, and the three- and two-dimensional cases with
projectors of any dimensionality, including degenerate operators, i.e., zero and unity. The
eight-dimensional calculation covers the seven-dimensional case with four- and three-dimensional
projectors, the six-dimensional one with four-, three- and two-dimensional projectors, the
five-dimensional one with four-, three-, two- and one-dimensional projectors, and the four-, three-
and two-dimensional ones with any projectors.

The Schmidt decomposition of the maximally violating state vector
tells which subspaces of the component spaces it occupies. The
dimensionality of those subspaces is the Schmidt number, the
number of nonzero Schmidt coefficients. If this is less than the
number of dimensions of the component spaces, and if each
measurement operator projects the corresponding subspace onto
itself, then the orthogonal subspace may be dropped altogether,
reducing the dimensionality of the solution. The latter condition
is fulfilled if the matrix of each measurement operator is
block-diagonal in the Schmidt basis, such that all matrix elements
connecting the subspace the state vector occupies with the
orthogonal subspace are zero. Although this was not always true
for the results we got from the optimization procedure, it turned
out that we could always get rid of the unwanted nonzero matrix
elements without changing the optimum value achieved by penalizing
them in a subsequent optimization. In these cases there were
equally good solutions with different choices of the measurement
operators concerned. Thus a Schmidt number smaller than the
dimensionality of the component Hilbert space we considered always
led to a solution in a smaller space.

The number of dimensions of a measurement operator in the reduced
space is given by the dimensionality of the overlap of the
subspace it projects to with the subspace the state vector
occupies. We can easily get it as the trace of the corresponding
block of the matrix of the operator. The dimensionality of the
overlap of a five-dimensional and a three-dimensional subspace of
a six-dimensional space is either two or three. This is the reason
our six-dimensional calculation with three-dimensional projectors
may lead to a five-dimensional solution with both two- and
three-dimensional measurement operators. To try all these
possibilities for a Bell inequality with five measurement settings
per party with confining ourselves to five-dimensional component
spaces would mean $2^{10}$ separate optimizations. To check all
potential four-dimensional solutions a single six-dimensional
calculation covers would require $3^{10}$ complete calculations.
Although the number of parameters to be optimized is larger, it is
still better to work in higher-dimensional spaces.

In our previous work \cite{PV08} we studied many of the Bell inequalities we consider here, but then
we confined ourselves to four-dimensional component spaces with two-dimensional projectors. We solved
the problem by maximizing the maximum eigenvalue of the $16\times 16$ matrix of the operator given by
Eq.~(\ref{Bellquant}) with a simplex uphill method. A difficulty with this approach is that it almost
always finds a solution with qubits, even if a better solution exists with higher dimensional
component spaces. For some inequalities it finds the true maximum only once in several thousand
attempts. With four-dimensional spaces the number of calculations necessary to get reliable results
were well affordable. However, in six- or eight-dimensional component spaces it is even more difficult
to find the interesting solution, there are a lot more parameters to be optimized, and the size of the
matrices to be diagonalized many times in each optimization run is also much larger. Therefore, we
have changed our strategy and we have used an iterative approach. The procedure ensures that the
Schmidt bases of the state vector agree with the bases used in the component spaces, up to possible
phase factors.

When we try to find the best $m$-dimensional solution with the $n$-dimensional ($n=6$ or $8$) program,
we first take the $m$-dimensional maximally entangled state
$1/\sqrt{m}\sum_{i=1,m}|i\rangle|i\rangle$. We apply the simplex uphill method \cite{NM65} to find the
best measurement operators to maximize the expectation value of the operator (\ref{Bellquant}), that
is the quantum value of the inequality with the state chosen. Then we replace the vector with the one
of the form $\sum_{i=1,n}c_i|i\rangle|i\rangle$ that gives the largest violation with the measurement
operators we have got. This vector can directly be determined by finding the eigenvector belonging to
the largest eigenvalue of a matrix. We repeat these steps until we reach convergence. As the vector
always has the special form in our basis, the matrices whose expectation value has to be calculated
and the ones to be diagonalized are of the size of $n\times n$. We only have to deal with small
submatrices of the corresponding matrices of size $n^2\times n^2$ in the full Hilbert space. For the
simplex uphill method we use the implementation of Ref.~\cite{numrec}. The procedure does not ensure
to find the true optimum, so we have to repeat it many times. To save computation time we made just a
few iterations with many different initial values, and chose only a few of them that arrived at the
best values for the full calculation. In another implementation we stopped the iterations as soon as
any of the Schmidt coefficients (which are just $|c_i|$) fell below a threshold, and restarted the
procedure with new initial values.

To apply optimization we have to parametrize the operators. An
$(n/2)$-dimensional projector in an $n$-dimensional space requires
$(n/2)^2$ real parameters to characterize in the real case, and
twice as many in the complex case. In choosing the parameters we
used a similar approach than in Ref.~\cite{PV08}. In
Ref.~\cite{TV06} it has been shown that any $(n/2)$-dimensional
projector in an $n$-dimensional space can be transformed into a
simple form with two unitary transformations, one acting only on
the first $n/2$, while the other on the other $n/2$ basis vectors.
This simple form is characterized by $n/2$ angles. Then any such
projector can be produced by acting on such a simple matrix with
two appropriate unitary --- or in the real case orthogonal ---
transformations. A general three-dimensional orthogonal transformation
can be described with 3 Euler angles, while the four-dimensional one
requires 6 Euler angles (see for example Ref.~\cite{Vilenkin}).
This way we get the correct number of parameters for the
projectors in the real case both in the six- $(3+2\cdot 3=9)$ and
in the eight-dimensional $(4+2\cdot 6=16)$ spaces. In a
six-dimensional space each three-dimensional unitary transformation
requires further 4 phase factors to be characterized, but one
overall factor is irrelevant. The 9 extra parameters we get this
way give just the necessary number of parameters for the
projectors in the complex case.

\section{Upper bounds}\label{upper}

\begin{table*}[tbm]
 \caption{Maximum violation of Bell inequalities. The identitity $\cal H$ of the
 component Hilbert spaces, the dimensionalities $D_{A_i}$ and $D_{B_j}$ of the
 measurement operators of the participants, and the level L when the upper
 limit reaches the value of violation is also shown. Notation 1a, 2a and 2b
 mean level 1 plus $ab$, level 2 plus $aa'b$ (or $abb'$) and level 2 plus
 $aa'b$+$abb'$, respectively. An alternative notation for $I_{3322}$,
 $I^2_{3422}$ and $I^1_{4422}$ is $A_{3}$, $A_{4}$ and $A_{7}$, respectively.}
 \vskip 0.2truecm
 \centering
 \begin{tabular}{l l l l l l l l l l l l l l l l l l}
 \hline\hline
 &Violation&$\cal H$&$D_{A_i}$&$D_{B_j}$&L&&Violation&
 $\cal H$&$D_{A_i}$&$D_{B_j}$&L&&Violation&$\cal H$&$D_{A_i}$&$D_{B_j}$&L\\
 \hline
 $I_{3322}$&0.2500000&$\RR^2$&111&111&no&\hphantom{v}$A_{16}$&0.4571068&$\CC^2$&0111&11111&2a&\hphantom{v}$A_{53}$&0.6386102&$\RR^2$&11111&11111&2a\\
 $I^1_{3422}$&0.4142136&$\RR^2$&011&1111&1a&\hphantom{v}$A_{17}$&0.3754473&$\RR^2$&1111&01111&2&\hphantom{v}$A_{54}$&0.5936813&$\RR^2$&11111&11111&2a\\
 $I^2_{3422}$&0.2990381&$\RR^2$&111&0111&2a&\hphantom{v}$A_{18}$&0.3843551&$\RR^2$&1111&01111&1a&\hphantom{v}$A_{55}$&0.6213203&$\RR^2$&01111&11111&1a\\
 $I^3_{3422}$&0.4364917&$\RR^2$&111&1111&1a&\hphantom{v}$A_{19}$&0.6226300&$\CC^2$&1111&11111&2a&\hphantom{v}$A_{56}$&0.6893124&$\RR^2$&11201&11111&2a\\
 $A_5$&0.4353342&$\RR^2$&1111&1111&2a&\hphantom{v}$A_{20}$&0.6022398&$\CC^2$&1111&11111&2a&\hphantom{v}$A_{57}$&0.6603444&$\RR^2$&11111&11111&2a\\
 $A_6$&0.3003638&$\RR^4$&1222&1222&3&\hphantom{v}$A_{21}$&0.3258362&$\RR^5$&2242&22323&no&\hphantom{v}$A_{58}$&0.6488905&$\RR^2$&11111&11111&2a\\
 $AS_1$&0.5412415&$\RR^2$&1111&1111&1&\hphantom{v}$A_{22}$&0.6234571&$\RR^2$&1111&11111&2a&\hphantom{v}$A_{59}$&0.4488256&$\RR^2$&11111&01111&2a\\
 $AS_2$&0.8784928&$\RR^2$&1111&1111&1&\hphantom{v}$A_{23}$&0.5460735&$\CC^2$&11111&11111&2&\hphantom{v}$A_{60}$&0.3940032&$\RR^5$&22242&22243&2b\\
 $AII_1$&0.6055543&$\RR^2$&1111&1111&1a&\hphantom{v}$A_{24}$&0.6047986&$\RR^2$&11111&11111&1a&\hphantom{v}$A_{61}$&0.4019248&$\CC^2$&11111&01111&2a\\
 $AII_2$&0.5000000&$\RR^2$&1111&1111&2&\hphantom{v}$A_{25}$&0.6033789&$\RR^2$&11111&11111&1a&\hphantom{v}$A_{62}$&0.4043897&$\RR^5$&22332&13233&no\\
 $I^1_{4422}$&0.2878683&$\RR^3$&1111&1111&2a&\hphantom{v}$A_{26}$&0.5275550&$\CC^2$&11111&11111&2&\hphantom{v}$A_{63}$&0.4894164&$\RR^4$&11222&11222&2a\\
 $I^2_{4422}$&0.6213712&$\RR^2$&1111&1111&2&\hphantom{v}$A_{27}$&0.6483073&$\RR^2$&11111&11111&2&\hphantom{v}$A_{64}$&0.3900890&$\RR^3$&12122&11222&no\\
 $I^3_{4422}$&0.4142136&$\RR^2$&2111&1111&1a&\hphantom{v}$A_{28}$&0.6403143&$\RR^2$&11111&11111&1a&\hphantom{v}$A_{65}$&0.3688702&$\RR^5$&12333&12333&no\\
 $I^4_{4422}$&0.4142136&$\RR^2$&1100&1111&2a&\hphantom{v}$A_{29}$&0.4920635&$\CC^2$&11111&11111&2&\hphantom{v}$A_{66}$&0.4877093&$\CC^2$&01111&11111&2a\\
 $I^5_{4422}$&0.4364917&$\RR^2$&1111&1111&1a&\hphantom{v}$A_{30}$&0.5698209&$\RR^2$&11111&11111&1a&\hphantom{v}$A_{67}$&0.3990671&$\RR^5$&22333&23332&no\\
 $I^6_{4422}$&0.4494897&$\CC^2$&1111&1111&1a&\hphantom{v}$A_{31}$&0.5738173&$\RR^2$&11111&11111&1a&\hphantom{v}$A_{68}$&0.4011462&$\RR^5$&22333&23332&no\\
 $I^7_{4422}$&0.4548373&$\CC^2$&1111&1111&2&\hphantom{v}$A_{32}$&0.4135530&$\CC^2$&11111&11111&2a&\hphantom{v}$A_{69}$&0.6096103&$\RR^2$&01111&11111&2a\\
 $I^8_{4422}$&0.4877681&$\RR^3$&2121&2112&2b&\hphantom{v}$A_{33}$&0.6226313&$\CC^2$&11111&11111&2a&\hphantom{v}$A_{70}$&0.6052228&$\RR^2$&01111&11111&2\\
 $I^9_{4422}$&0.4616842&$\RR^2$&1111&1111&2a&\hphantom{v}$A_{34}$&0.5350117&$\RR^2$&11211&11111&2a&\hphantom{v}$A_{71}$&0.4490163&$\RR^2$&01111&11111&2a\\
 $I^{10}_{4422}$&0.6139456&$\RR^2$&1111&1111&2a&\hphantom{v}$A_{35}$&0.6249079&$\RR^2$&11111&11111&2a&\hphantom{v}$A_{72}$&0.6962822&$\RR^2$&11111&11111&2a\\
 $I^{11}_{4422}$&0.6383543&$\RR^2$&1111&1111&1a&\hphantom{v}$A_{36}$&0.4388685&$\CC^2$&11111&11111&2a&\hphantom{v}$A_{73}$&0.8831381&$\RR^2$&11111&11011&2a\\
 $I^{12}_{4422}$&0.6188142&$\RR^2$&1111&1111&2a&\hphantom{v}$A_{37}$&0.4868868&$\CC^2$&11111&11111&2a&\hphantom{v}$A_{74}$&0.6890694&$\RR^2$&11111&11111&2a\\
 $I^{13}_{4422}$&0.4348553&$\RR^2$&1112&1111&2a&\hphantom{v}$A_{38}$&0.4699126&$\CC^2$&11111&11111&2a&\hphantom{v}$A_{75}$&0.6051510&$\RR^2$&11111&11011&2a\\
 $I^{14}_{4422}$&0.4794103&$\RR^2$&1110&1110&2a&\hphantom{v}$A_{39}$&0.6172035&$\CC^2$&11111&11111&2&\hphantom{v}$A_{76}$&0.4898631&$\RR^3$&11111&11111&2a\\
 $I^{15}_{4422}$&0.4348553&$\RR^2$&1011&1111&2a&\hphantom{v}$A_{40}$&0.6078638&$\RR^2$&11111&11111&1a&\hphantom{v}$A_{77}$&0.6655582&$\RR^2$&11111&11111&2a\\
 $I^{16}_{4422}$&0.4142136&$\RR^2$&1011&1011&1a&\hphantom{v}$A_{41}$&0.4785634&$\CC^2$&11111&11111&2a&\hphantom{v}$A_{78}$&0.8927018&$\RR^2$&11111&11111&2\\
 $I^{17}_{4422}$&0.6714085&$\RR^2$&1111&1111&2a&\hphantom{v}$A_{42}$&0.6198655&$\RR^2$&11111&11111&1a&\hphantom{v}$A_{79}$&0.6243153&$\CC^2$&11111&11111&2a\\
 $I^{18}_{4422}$&0.6429670&$\RR^3$&1111&1111&2a&\hphantom{v}$A_{43}$&0.6107654&$\RR^2$&11111&11111&1a&\hphantom{v}$A_{80}$&0.3769863&$\RR^4$&12212&21232&no\\
 $I^{19}_{4422}$&0.4971707&$\RR^3$&2112&2122&2b&\hphantom{v}$A_{44}$&0.5364942&$\RR^2$&11211&11111&2a&\hphantom{v}$A_{81}$&0.6690099&$\CC^2$&11111&11111&2a\\
 $I^{20}_{4422}$&0.4676794&$\RR^4$&2223&2223&no&\hphantom{v}$A_{45}$&0.5372394&$\CC^2$&11111&10111&2a&\hphantom{v}$A_{82}$&0.4644255&$\RR^4$&21222&11222&no\\
 $A_8$&0.5916501&$\CC^2$&1111&11111&1a&\hphantom{v}$A_{46}$&0.4590108&$\RR^5$&23333&22233&2b&\hphantom{v}$A_{83}$&0.6961664&$\CC^2$&11111&11111&2a\\
 $A_9$&0.4652428&$\CC^2$&1111&11111&2a&\hphantom{v}$A_{47}$&0.4608544&$\CC^2$&11111&11111&no&\hphantom{v}$A_{84}$&0.6332967&$\RR^4$&23222&22212&no\\
 $A_{10}$&0.4158004&$\RR^5$&3233&23333&2b&\hphantom{v}$A_{48}$&0.4631707&$\RR^4$&11222&22221&2a&\hphantom{v}$A_{85}$&0.6411408&$\CC^2$&11111&11111&2a\\
 $A_{11}$&0.4561079&$\CC^2$&1111&11111&2a&\hphantom{v}$A_{49}$&0.4666943&$\CC^2$&11111&11111&2a&\hphantom{v}$A_{86}$&0.8004425&$\CC^2$&11111&11111&2a\\
 $A_{12}$&0.4877093&$\CC^2$&1111&11111&2&\hphantom{v}$A_{50}$&0.5182900&$\CC^2$&11111&11111&2a&\hphantom{v}$A_{87}$&0.7562471&$\RR^4$&11111&11111&2a\\
 $A_{13}$&0.4252330&$\RR^4$&1222&12222&2a&\hphantom{v}$A_{51}$&0.6607809&$\RR^2$&11111&11111&2a&\hphantom{v}$A_{88}$&0.4142136&$\RR^2$&00111&11111&2\\
 $A_{14}$&0.4758457&$\RR^5$&3322&22232&no&\hphantom{v}$A_{52}$&0.6218611&$\RR^2$&11111&11111&2a&\hphantom{v}$A_{89}$&0.3025898&$\RR^6$&22343&22343&no\\
 $A_{15}$&0.4496279&$\CC^2$&1111&11111&2a\\
 \hline
 \end{tabular}
 \label{table:b5}
 \end{table*}

To determine upper limits for the maximum quantum violation of the Bell inequalities we follow the
procedure given by Navascu\'{e}s et al. \cite{NPA07,NPA08}. The method, in its simplest form,
specified to the case of a two-party two-outcome Bell inequality, is as follows.

Let us introduce the operators $a_i=(2A_i-I_A)\otimes I_B$ and $b_j=I_A\otimes (2B_j-I_B)$, where
$I_A$ and $I_B$ are the unity operators in Alice's and Bob's component spaces, respectively. The $a_i$
and $b_j$ operators have eigenvalues $+1$ and $-1$ for the two outcomes of the measurement instead of
$1$ and $0$. The square of each of them is unity, and every $a_i$ commutes with every $b_j$. Let us
choose a set of vectors containing $|\psi\rangle$, $a_1|\psi\rangle$, $a_2|\psi\rangle,\dots,
a_{m_A}|\psi\rangle$, $b_1|\psi\rangle$, $b_2|\psi\rangle,\dots, b_{m_B}|\psi\rangle$, and any number
of vectors of the form $a_ia_{i'}a_{i''}...b_jb_{j'}b_{j''}|\psi\rangle$, i.e., $|\psi\rangle$
multiplied by the product of any numbers of $a$ and $b$ operators. Let us consider the matrix whose
entries are the real parts of the scalar products of these vectors. A matrix derived this way is
positive semidefinite. Each entry of the matrix is the real part of the expectation value of an
operator. Taking into account that each $a_i$ and $b_j$ squares to unity, and that $a_i$ commutes with
$b_j$, one can see that in some places the operators may be equal, or hermitian conjugates of each
other. In those places the entries of the matrix will be equal, independently of the actual choice of
$|\psi\rangle$ and the measurement operators. It is easy to see that the quantum value for a Bell
inequality can be written as a linear combination of certain entries of this matrix. There are limits
on the value this linear combination may take, as the matrix has to be positive semidefinite, while
satisfying constraints requiring the equality of some of its elements. The quantum value of the Bell
inequality can not be larger than the maximum value allowed by the conditions above. The standard
optimization method to determine this maximum is semidefinite programming \cite{VB96}. There are
several codes available to solve the problem, in most of our calculations we used CSDP \cite{CSDP}.

The upper limit for the maximum violation of the inequality we get this way depends on how we choose
the set of vectors we generate the matrix from. Refs.~\cite{NPA07,NPA08} introduce a systematic
hierarchy of choices, where the higher levels give ever stricter limits until the exact maximum
violation is achieved. The first level includes only the vectors  $|\psi\rangle$, $a_i|\psi\rangle$
($i=1,\dots,m_A$) and $b_j|\psi\rangle$ ($j=1,\dots,m_B$). The second level also includes vectors
$a_ia_{i'}|\psi\rangle$, $b_jb_{j'}|\psi\rangle$ and $a_ib_j|\psi\rangle$ ($i,i'=1,\dots,m_A$,
$j,j'=1,\dots,m_B$), that is it includes all vectors we can make by acting on $|\psi\rangle$ with up
to two measurement operators. We get level three by going up to three operators, and so on. The exact
value for the maximum violation may be reached at a finite level, for correlation type Bell
inequalities this already happens at the first level \cite{Wehner06}. It has been shown that the
series of upper limits through the hierarchy converges to the exact value \cite{NPA08,DLTW08}. In
practice, it is also worth considering intermediate levels. The computation requirement is often much
less than that is for the next full level, and if it gives the exact value, we need not go any
further. For example, for many of the Bell inequalities we considered we got the upper limit equal to
the lower one by supplementing level two by just the vectors generated with two $a$ and one $b$ (noted
as $aa'b$), or one $a$ and two $b$ operators ($abb'$).

\section{Results for the maximum violation}\label{disc}

 \begin{table*}[p]
 \caption{Maximum violation of Bell inequalities introduced in the present paper
 The identitity $\cal H$ of the
 component Hilbert spaces, the dimensionalities $D_{A_i}$ and $D_{B_j}$ of the
 measurement operators of the participants, and the level L when the upper
 limit reaches the value of violation is also shown. Notation 1a, 2a and 2b
 mean level 1 plus $ab$, level 2 plus $aa'b$ (or $abb'$) and level 2 plus
 $aa'b$+$abb'$, respectively.}
 \vskip 0.2truecm
 \centering
 \begin{tabular}{l l l l l l l l l l l l l l l l l l}
 \hline\hline
 &Violation&$\cal H$&$D_{A_i}$&$D_{B_j}$&L&&Violation&
 $\cal H$&$D_{A_i}$&$D_{B_j}$&L&&Violation&$\cal H$&$D_{A_i}$&$D_{B_j}$&L\\
 \hline
 $J^{1}_{4422}$&0.4554438&$\RR^3$&2122&1111&2a&\hphantom{v}$J^{44}_{4422}$&0.4144533&$\RR^2$&1111&1111&2a&\hphantom{v}$J^{87}_{4422}$&0.6848726&$\RR^2$&1111&1111&2a\\
 $J^{2}_{4422}$&0.6140029&$\RR^2$&1111&1111&2a&\hphantom{v}$J^{45}_{4422}$&0.7705600&$\RR^2$&1121&1110&2a&\hphantom{v}$J^{88}_{4422}$&0.6159879&$\RR^2$&1111&1111&2\\
 $J^{3}_{4422}$&0.8106306&$\RR^2$&1111&1211&2a&\hphantom{v}$J^{46}_{4422}$&0.9716836&$\CC^2$&1111&1111&2a&\hphantom{v}$J^{89}_{4422}$&1.0035502&$\RR^2$&1111&1111&1a\\
 $J^{4}_{4422}$&0.6861985&$\CC^2$&1111&1111&2a&\hphantom{v}$J^{47}_{4422}$&0.7644749&$\RR^2$&1111&1111&2a&\hphantom{v}$J^{90}_{4422}$&0.8398157&$\RR^2$&1111&1111&1a\\
 $J^{5}_{4422}$&0.6520359&$\RR^2$&1011&1111&2a&\hphantom{v}$J^{48}_{4422}$&0.7510516&$\RR^6$&3334&3433&no&\hphantom{v}$J^{91}_{4422}$&1.2992769&$\RR^2$&1111&1111&1a\\
 $J^{6}_{4422}$&0.4461508&$\RR^2$&1011&1111&2a&\hphantom{v}$J^{49}_{4422}$&0.8156923&$\RR^4$&3222&2232&2a&\hphantom{v}$J^{92}_{4422}$&1.0648162&$\RR^2$&1111&1111&2a\\
 $J^{7}_{4422}$&0.6057263&$\RR^2$&1111&1211&2a&\hphantom{v}$J^{50}_{4422}$&0.8555932&$\RR^2$&1111&1111&1a&\hphantom{v}$J^{93}_{4422}$&0.9627261&$\RR^2$&1111&1111&1a\\
 $J^{8}_{4422}$&0.4772413&$\RR^2$&1211&1121&2a&\hphantom{v}$J^{51}_{4422}$&0.6750341&$\RR^2$&1121&1111&2a&\hphantom{v}$J^{94}_{4422}$&0.5944550&$\RR^2$&0111&1111&2a\\
 $J^{9}_{4422}$&0.5000000&$\RR^2$&1111&1011&2&\hphantom{v}$J^{52}_{4422}$&1.0999033&$\RR^2$&1111&1111&1a&\hphantom{v}$J^{95}_{4422}$&1.0014114&$\RR^2$&0111&1111&2a\\
 $J^{10}_{4422}$&0.6938425&$\RR^3$&2222&2121&2b&\hphantom{v}$J^{53}_{4422}$&0.8093253&$\RR^2$&0111&1101&2a&\hphantom{v}$J^{96}_{4422}$&0.9416515&$\CC^2$&1111&1111&2a\\
 $J^{11}_{4422}$&0.4684758&$\RR^4$&2222&2121&2b&\hphantom{v}$J^{54}_{4422}$&0.5194497&$\RR^3$&2121&1221&2a&\hphantom{v}$J^{97}_{4422}$&1.1583626&$\RR^2$&1111&1111&no\\
 $J^{12}_{4422}$&0.7261971&$\RR^4$&2322&2322&2b&\hphantom{v}$J^{55}_{4422}$&0.8283881&$\RR^2$&2111&1121&2a&\hphantom{v}$J^{98}_{4422}$&1.1496911&$\RR^2$&1111&1111&1a\\
 $J^{13}_{4422}$&0.6926870&$\RR^3$&1121&1111&2b&\hphantom{v}$J^{56}_{4422}$&0.8192037&$\RR^3$&2111&1121&2a&\hphantom{v}$J^{99}_{4422}$&0.8990726&$\RR^2$&0111&1111&2a\\
 $J^{14}_{4422}$&0.6317469&$\RR^2$&1111&1211&2a&\hphantom{v}$J^{57}_{4422}$&0.8609885&$\RR^2$&0111&1101&2a&\hphantom{v}$J^{100}_{4422}$&1.2675376&$\RR^2$&1111&1111&2\\
 $J^{15}_{4422}$&0.7808500&$\RR^2$&1111&1211&2&\hphantom{v}$J^{58}_{4422}$&0.8814481&$\RR^2$&1111&1111&1a&\hphantom{v}$J^{101}_{4422}$&1.0296105&$\RR^2$&1111&1111&2a\\
 $J^{16}_{4422}$&0.5674133&$\RR^2$&1111&1111&2a&\hphantom{v}$J^{59}_{4422}$&0.6379629&$\RR^2$&0111&1101&2&\hphantom{v}$J^{102}_{4422}$&0.6650865&$\RR^2$&1111&1111&2\\
 $J^{17}_{4422}$&0.6379629&$\RR^2$&2111&1121&2a&\hphantom{v}$J^{60}_{4422}$&0.5922714&$\RR^2$&1111&1111&1a&\hphantom{v}$J^{103}_{4422}$&1.0520136&$\RR^2$&1111&1111&2\\
 $J^{18}_{4422}$&1.0130080&$\RR^2$&1101&1101&1a&\hphantom{v}$J^{61}_{4422}$&0.8175538&$\RR^2$&1111&1111&2a&\hphantom{v}$J^{104}_{4422}$&1.5875885&$\RR^2$&1111&1111&1a\\
 $J^{19}_{4422}$&0.6742346&$\RR^2$&1111&1111&1a&\hphantom{v}$J^{62}_{4422}$&0.7500754&$\RR^6$&3333&3333&no&\hphantom{v}$J^{105}_{4422}$&1.0742228&$\RR^2$&1111&1111&1a\\
 $J^{20}_{4422}$&0.5932470&$\RR^2$&1110&1121&2a&\hphantom{v}$J^{63}_{4422}$&0.6078543&$\RR^4$&2222&2212&2a&\hphantom{v}$J^{106}_{4422}$&0.8382498&$\RR^2$&1111&1111&1a\\
 $J^{21}_{4422}$&0.7017468&$\RR^3$&2221&2212&2a&\hphantom{v}$J^{64}_{4422}$&0.9167082&$\CC^2$&1111&1111&2a&\hphantom{v}$J^{107}_{4422}$&0.9339703&$\RR^2$&1111&1111&1a\\
 $J^{22}_{4422}$&0.8156179&$\RR^2$&1111&1111&1a&\hphantom{v}$J^{65}_{4422}$&1.1110314&$\RR^2$&1111&1111&2&\hphantom{v}$J^{108}_{4422}$&0.9676411&$\RR^2$&1111&1111&1a\\
 $J^{23}_{4422}$&0.6090289&$\CC^2$&1111&1111&2a&\hphantom{v}$J^{66}_{4422}$&0.6186278&$\RR^2$&1111&1111&2a&\hphantom{v}$J^{109}_{4422}$&1.7260512&$\RR^2$&1111&1111&1a\\
 $J^{24}_{4422}$&0.6942505&$\RR^2$&1111&1111&2a&\hphantom{v}$J^{67}_{4422}$&0.7212948&$\RR^2$&1111&1110&2a&\hphantom{v}$J^{110}_{4422}$&0.9457465&$\RR^2$&1111&1111&1a\\
 $J^{25}_{4422}$&0.5147829&$\RR^2$&1111&1211&2a&\hphantom{v}$J^{68}_{4422}$&1.0178506&$\RR^2$&1111&1111&1a&\hphantom{v}$J^{111}_{4422}$&0.7575890&$\RR^2$&1111&1111&1a\\
 $J^{26}_{4422}$&0.6402494&$\RR^2$&1111&1111&2a&\hphantom{v}$J^{69}_{4422}$&0.5081419&$\RR^2$&1111&1211&2a&\hphantom{v}$J^{112}_{4422}$&0.6247585&$\RR^2$&1111&1111&2a\\
 $J^{27}_{4422}$&0.9642857&$\CC^2$&1111&1111&2&\hphantom{v}$J^{70}_{4422}$&0.9538824&$\RR^2$&1111&1111&1a&\hphantom{v}$J^{113}_{4422}$&0.8483764&$\RR^2$&1112&1111&3\\
 $J^{28}_{4422}$&0.7500000&$\RR^2$&1111&1111&2a&\hphantom{v}$J^{71}_{4422}$&0.5822872&$\RR^2$&1111&1111&2a&\hphantom{v}$J^{114}_{4422}$&1.0653155&$\CC^2$&1111&1111&2a\\
 $J^{29}_{4422}$&1.0245760&$\CC^2$&1111&1111&2a&\hphantom{v}$J^{72}_{4422}$&0.7878251&$\CC^2$&1111&1111&2a&\hphantom{v}$J^{115}_{4422}$&1.0098044&$\RR^2$&1111&1101&2a\\
 $J^{30}_{4422}$&0.4361950&$\RR^5$&3231&2323&no&\hphantom{v}$J^{73}_{4422}$&1.1204592&$\CC^2$&1111&1111&no&\hphantom{v}$J^{116}_{4422}$&0.7651941&$\CC^2$&1111&1111&2a\\
 $J^{31}_{4422}$&0.4772413&$\RR^2$&1211&1121&2a&\hphantom{v}$J^{74}_{4422}$&0.4348553&$\RR^2$&2111&1111&2a&\hphantom{v}$J^{117}_{4422}$&0.9721185&$\CC^2$&1111&1111&2a\\
 $J^{32}_{4422}$&0.5901592&$\RR^2$&1111&1111&2a&\hphantom{v}$J^{75}_{4422}$&0.6270262&$\RR^2$&1111&1111&1a&\hphantom{v}$J^{118}_{4422}$&0.5283248&$\CC^2$&1111&1111&2a\\
 $J^{33}_{4422}$&0.6150642&$\RR^2$&1111&1111&2a&\hphantom{v}$J^{76}_{4422}$&0.8291077&$\RR^2$&1111&1111&1a&\hphantom{v}$J^{119}_{4422}$&0.8986088&$\RR^2$&1111&1112&2a\\
 $J^{34}_{4422}$&0.4492657&$\RR^4$&2222&2312&no&\hphantom{v}$J^{77}_{4422}$&0.8406242&$\RR^2$&1111&1111&2&\hphantom{v}$J^{120}_{4422}$&0.9333181&$\RR^2$&1121&1111&2a\\
 $J^{35}_{4422}$&0.6932503&$\RR^2$&1111&1111&2a&\hphantom{v}$J^{78}_{4422}$&0.8937709&$\RR^2$&1111&1110&2a&\hphantom{v}$J^{121}_{4422}$&0.5970802&$\RR^2$&1111&1111&2a\\
 $J^{36}_{4422}$&0.6706462&$\RR^2$&1111&1111&2a&\hphantom{v}$J^{79}_{4422}$&1.0156152&$\RR^2$&1111&1111&2a&\hphantom{v}$J^{122}_{4422}$&0.7931031&$\RR^2$&1111&1101&2\\
 $J^{37}_{4422}$&0.4864826&$\RR^3$&1121&1112&2a&\hphantom{v}$J^{80}_{4422}$&0.9051257&$\RR^2$&1111&1111&2&\hphantom{v}$J^{123}_{4422}$&1.0640930&$\CC^2$&1111&1111&2a\\
 $J^{38}_{4422}$&0.9295153&$\RR^2$&1111&1111&2a&\hphantom{v}$J^{81}_{4422}$&0.5995871&$\RR^2$&1111&1111&2&\hphantom{v}$J^{124}_{4422}$&0.9409648&$\RR^2$&1111&1111&2\\
 $J^{39}_{4422}$&0.8264535&$\RR^2$&1111&1111&2a&\hphantom{v}$J^{82}_{4422}$&0.7308078&$\CC^2$&1111&1111&2a&\hphantom{v}$J^{125}_{4422}$&1.0000000&$\RR^2$&1111&1111&2a\\
 $J^{40}_{4422}$&0.9208907&$\RR^4$&2212&2221&3&\hphantom{v}$J^{83}_{4422}$&0.6691786&$\RR^2$&1111&1111&2a&\hphantom{v}$J^{126}_{4422}$&0.8023577&$\CC^2$&1111&1111&2a\\
 $J^{41}_{4422}$&0.7596044&$\RR^2$&1111&1111&2&\hphantom{v}$J^{84}_{4422}$&1.0669560&$\RR^2$&1111&1111&2a&\hphantom{v}$J^{127}_{4422}$&0.6718972&$\RR^3$&1222&2121&2a\\
 $J^{42}_{4422}$&0.6722257&$\RR^2$&1111&1121&2a&\hphantom{v}$J^{85}_{4422}$&0.9762793&$\RR^2$&1111&1111&2&\hphantom{v}$J^{128}_{4422}$&1.0096066&$\RR^2$&1111&1111&1a\\
 $J^{43}_{4422}$&0.4685939&$\RR^3$&2221&2212&2a&\hphantom{v}$J^{86}_{4422}$&0.7558762&$\RR^2$&1112&1111&2b&\hphantom{v}$J^{129}_{4422}$&1.0522569&$\RR^2$&1111&1111&1a\\
 \hline
 \end{tabular}
 \label{table:vt}
 \end{table*}

 \begin{table*}[t]
 \caption{Difference between the upper limit calculated at different levels
 and the lower limit derived by explicitly constructing the measurement operators.}
 \vskip 0.2truecm
 \centering
 \begin{tabular}{l c c c c c c c}
 \hline\hline

Case\hphantom{W}&$L2$&\hphantom{+}$L2+aa'b$\hphantom{+}&$L2+85\%\ \rm{of}$&$L2+aa'b+$&$L2+aa'b+$&$L3$&$L3+aa'a''b+$\\
&&&$aa'b+abb'$&$+abb'$&$abb'+aa'a''$&&$aa'bb'+abb'b''$\\
 \hline
$I_{3322}$&0.0009397&0.0008863&&0.0008760&0.0008756&0.0008756&0.0008754\\
$I^{20}_{4422}$&0.0393814&0.0111147&&0.0040799&&0.0000784\\
$J^{30}_{4422}$&0.0401921&0.0109109&&0.0042758&&0.0010680\\
$J^{34}_{4422}$&0.0240714&0.0064382&&0.0023265&&0.0001503\\
$J^{48}_{4422}$&0.0294613&0.0014661&&0.0008068&&0.0006361\\
$J^{62}_{4422}$&0.0022695&0.0000194&&0.0000178&&0.0000167\\
$J^{73}_{4422}$&0.0329957&0.0030318&&0.0023042&&0.0020987\\
$J^{97}_{4422}$&0.0392054&0.0018759&&0.0005867&&0.0001425\\
$A_{14}$&0.0274655&0.0057079&&0.0031815&0.0025615\\
$A_{21}$&0.0048537&0.0007595&&0.0005165&0.0003840\\
$A_{47}$&0.0202626&0.0050055&0.0026688\\
$A_{62}$&0.0133913&0.0031907&0.0025556\\
$A_{64}$&0.0164779&0.0032730&0.0016572\\
$A_{65}$&0.0175447&0.0014105&0.0000806\\
$A_{67}$&0.0135307&0.0019890&0.0006118\\
$A_{68}$&0.0073408&0.0039257&0.0038884\\
$A_{80}$&0.0161885&0.0026707&0.0006052\\
$A_{82}$&0.0082846&0.0065568&0.0064928\\
$A_{84}$&0.0271831&0.0040203&0.0021524\\
$A_{89}$&0.0104965&0.0043258&0.0033485\\
 \hline
 \end{tabular}
 \label{table:highlowdiff}
 \end{table*}

Tables~\ref{table:b5} and \ref{table:vt} show our results for the maximum violation of the Bell
inequalities we could achieve by explicitly determining the state vector and the measurement
operators. The size of the component Hilbert space and the dimensionalities of the operators are also
shown. In our earlier work \cite{PV08} we have determined the maximum violation for all cases shown in
Table~\ref{table:b5} with component Hilbert spaces of up to four dimensions taking two-dimensional
projectors as measurement operators. For the inequalities we follow the notations we used there, which
are the same as the ones in Refs.~\cite{BrunGis,89list}. By extending our calculations to
six-dimensional complex and real, and eight-dimensional real component spaces we could achieve a
larger maximum violation for 19 of the 112 inequalities. We found 9 cases where four-dimensional real
spaces were necessary for maximum violation, with not all projectors two-dimensional. For another 9
inequalities we had to go up to five-dimensional real component spaces. One of them is $A_{10}$, a
4522 inequality, which has less than five measurement settings for one of the parties. For one 5522 we
found the maximum violation with six-dimensional component spaces, more than the number of measurement
settings for either party. Similarly, for $J^{30}_{4422}$ (see Table~\ref{table:vt}), we needed
five-dimensional, and for $J^{48}_{4422}$ and $J^{62}_{4422}$, all 4422 cases, six-dimensional real
component spaces. However, for these cases we can not be sure that the states and the operators we
constructed does give the true quantum bound. The values for the violation are still somewhat below
the corresponding upper limits we have got. Some possibilities, even with lower dimensional component
spaces, like four-dimensional complex spaces with allowing degenerate measurement operators are not
covered. Moreover, we may have even missed some solutions which are in principle accessible by our
programs. The six-dimensional program with complex spaces often struggles to find solutions it finds
relatively easily in real spaces. This is not very surprising, as in the complex case there are twice
as many parameters to optimize. We note that whenever we needed more than two-dimensional component
spaces for maximum violation for the inequalities considered in the present paper, real spaces were
always sufficient. This is not true for all Bell inequalities, for example for the 4822 correlation
type inequality introduced in Ref.~\cite{VP08} the lowest dimensional component spaces to achieve the
maximum violation are four-dimensional complex ones. Some complex solutions may have been missed by
our present calculations, but not many of them, as for the great majority of the inequalities the
solution we managed to construct are actually as good as they can be, according to the upper limit we
found.

For the cases the upper limit agreed with the lower one derived from the explicit construction, we
also show in Tables~\ref{table:b5} and \ref{table:vt} the (usually partial) level when it happened.
For those inequalities the maximum violations shown are the exact values. The level we could afford to
accomplish with the code and with the computers we used depended on the number of measurement
settings. For the 5522 inequalities we could do the calculation with either all $aa'b$ or $abb'$ type
vectors added to level two, but with not both of them. Therefore, the largest calculations we made
included only the $85\%$ of them, randomly chosen. For the 4522 cases the maximum we could do was
level two plus $aa'b+abb'+aa'a''$ (i.e., level 3 without $bb'b''$). For 4422 we could go up to level
3, while for 3322 to level three plus $aa'bb''+aa'a''b+abb'b''$. We had to go to the maximum we could
afford only for a small fraction of the inequalities. For over $85\%$ of the cases level two plus
$abb'$ (and for the same cases level two plus $aa'b$) has already given the exact value.

There were 20 inequalities for which the upper limit remained above the best solution we found (marked
by $no$ in the Tables) at all levels we could afford. The gap between the upper and lower limits for
those inequalities, depending on the level is shown in Table~\ref{table:highlowdiff}. In almost all of
these cases extending the calculation for the higher limit further decreases this gap, often quite
significantly. For $I^{20}_{4422}$, $J^{30}_{4422}$, $J^{34}_{4422}$, $J^{97}_{4422}$, $A_{65}$,
$A_{67}$ and $A_{80}$ it seems to be probable that the lower limits we got are exact. When the
decrease of the gap is relatively very small, like in the cases of $J^{73}_{4422}$, $A_{68}$ or
$A_{82}$, it is more likely that we still have not got the true solution.

However, we find the case of $I_{3322}$ truly puzzling. This is
one of the simplest tight Bell inequalities, with just three
measurement settings for each party. In our formulation used in
(\ref{Bell}) for the Bell inequalities it reads
\begin{align}
&-2p_{A_1}-p_{A_2}-p_{B_1}+p_{A_1B_1}+p_{A_1B_2}+p_{A_1B_3}\nonumber\\
&+p_{A_2B_1}+p_{A_2B_2}-p_{A_2B_3}+p_{A_3B_1}-p_{A_3B_2}\le 0.
\end{align}
We could calculate the upper limit at a level significantly
exceeding level 3 (up to level 3 the upper limit has also been
calculated by Refs.~\cite{NPA08,DLTW08}). The gap between the
lower limit of $0.25$ and the upper limit is at least 4 orders of
magnitude larger than numerical uncertainty, while it gets just
marginally smaller by the extension of the calculation for the
upper limit. It does not behave as if it would decrease much
further. The lower limit has already been achieved with real
two-dimensional component spaces. All our attempts to find a
larger violation with higher dimensional spaces failed, although
the number of parameters is much smaller than for the other
inequalities we considered, and we have also made a lot more
attempts than for any other case. It does not seem very likely for
us that with spaces and operators our calculations cover such a
solution exists. At the same time, it would also be very
surprising, if one needed even higher dimensional Hilbert spaces
to violate maximally this very simple inequality.

Our present results confirm what we have found in Ref.~\cite{PV08}, namely in most cases the state
giving the maximum violation is not the maximally entangled one (indication for this fact was also
given independently by Ref.~\cite{BrunGis}). The results here are stronger, because the comparison
with the upper bound proves that for most inequalities the states and operators we have got are
actually the ones giving the absolute quantum bound. When the component Hilbert spaces required were
larger than two-dimensional, the state we found was never the maximally entangled one. For bipartite
inequalities with two measurement settings, but more than two outcomes \cite{CGLMP}, a similar result
was found by Ref.~\cite{NPA07}.

\section{Detection efficiencies}\label{detect}

A loophole-free experiment in order to test the nonlocal nature of quantum mechanics is still missing
\cite{AS12}. In particular, none of the experiments performed to date could close simultaneously the
locality loophole (the measurement results at Alice's and Bob's side should be space-like separated)
and the detection loophole. In order to avoid this latter loophole the particles must be detected with
a high enough probability, otherwise a locally causal model can reproduce the measured correlations.
Another motivation beside the fundamental ones comes from the security issue of some quantum
communication protocols which is based on the loophole-free violation of Bell inequalities
\cite{qcrypt}.

There are different proposals to close the detection loophole. More than two outcomes \cite{MPRG,Big},
more than two settings \cite{MPRG,Big,BHMR,BrunGis,BGSS}, and partially entangled states
\cite{Eberhard,BGSS,CL07} has been considered as well. The case of asymmetric Bell experiments, where
the two particles are detected with different probabilities (in systems such as entangled atom-photon
pairs) has also been addressed \cite{BGSS,CL07,BrunGis}. Another promising approach is the application
of homodyne detectors in Bell tests \cite{Patron,PJR,ACFN}.

Here we calculate both the symmetric (Alice and Bob have the same efficiency
$\eta_A=\eta_B\equiv\eta$) and asymmetric (Alice's detector is perfect) threshold detection
efficiencies of our set of Bell inequalities for a pair of maximally entangled qubits allowing
degenerate measurements as well. The general approach, consisting of two different detection
efficiencies ($\eta_A$ and $\eta_B$) was treated in Ref.~\cite{BGSS}. According to their Eq.~(4) the
quantum value of a Bell inequality with detection efficiencies $\eta_A$ and $\eta_B$ is given by
\begin{align}
I_{\eta_A,\eta_B}&=\eta_A\eta_B Q + \eta_A(1-\eta_B)M_A\nonumber\\
&+ (1-\eta_A)\eta_B M_B +(1-\eta_A)(1-\eta_B)X,
\end{align}
where $Q$ is the quantum value associated to the Bell inequality with perfect detectors, $M_{A,B}$ and
$X$ are the values when one or both detectors do not fire. Next we limit ourselves to maximally
entangled states and in the case of no detection one of the two outcomes $\{1,0\}$ are taken as an
output. If we consider degenerate measurements as well, the no-detection outcomes must be set to the
same value as the output of the corresponding degenerate measurement. On the other hand, by fixing
these values the original Bell inequality $I$ reduces to another one $I'$ with a smaller number of
settings. The local bound $L$ for this inequality can be smaller or equal to the local bound
corresponding to the Bell inequality $I$. At the same time it can be observed that the values
$M_{A,B}$ and $X$ remain the same for $I'$. Thus, if for a given Bell inequality $I$ and efficiencies
$\eta_A$,$\eta_B$, $I_{\eta_A,\eta_B}>L$ holds, it certainly holds true for the reduced Bell
inequality $I'$ as well. This implies that the threshold efficiencies for $I'$ can only be lowered
with respect to the original one $I$ (in our actual calculations we found that $L$ never decreased,
hence leaving unchanged the detection efficiencies). Conversely, this result means that if one
supplements a Bell inequality with coefficients pertaining to degenerate measurements (i.e.,
deterministic ones, which need not be performed at all), it cannot lower the detection efficiency.
Note, that this result, which is intuitively clear, also holds for the case of partially entangled
states. We made an optimization over all possible measurement strategies (i.e., for any combination of
degenerate and non-degenerate measurements) and over all possible no-detection outcomes, both in the
symmetric and asymmetric situations. The following results have been obtained:

From the literature \cite{BGSS} for the maximally entangled, asymmetric case the best known detection
efficiency is $\eta_B=2/3$, corresponding to $I_{3322}$. We found that inequality $I^2_{3422}$ and
$AII_2$ yield the same value with non-degenerate measurements. With involving degenerate measurements
as well, this threshold value was obtained for a further 97 inequalities. Out of them 69 can be traced
back to $I_{3322}$, while 12 is reducible to non-tight 3322 inequalities, nonequivalent with
$I_{3322}$. However, $\eta_B=2/3$ is not the best one can achieve, for three five-setting Bell
inequalities from the list of Avis et al. \cite{89list}, namely, $A_{34}$, $A_{44}$, and $A_{50}$ do
better. The respective values of $\eta_B=0.6607$, $0.6520$, $0.6587$ are reached with purely
non-degenerate measurements. There exists a local model \cite{GG}, which reproduces the correlations
arising from non-degenerate projective measurements which can be expressed using real (complex)
numbers on the maximally entangled state under the assumption that Alice has perfect detector and Bob
has detection efficiency $\eta_B=2/3$ ($\eta_B=1/2$). Thus the three lower than $2/3$ values given
above need to belong to projective measurements with settings requiring complex numbers to be
described. This fact, as we have checked, is true.

For the maximally entangled symmetric case, the threshold efficiency for the CHSH inequality is known
to be $0.8284$, which result was slightly improved recently by Brunner and Gisin \cite{BrunGis}. They
showed that the $A_5$ inequality from the Avis et al.\ list allows the slightly smaller threshold of
$0.8214$. By computing the best threshold values for the set of 241 Bell inequalities we did not find
any better case. Allowing degenerate measurements we could get the same value for 19 five-setting Bell
inequalities, but they could all be traced back to $A_5$.

\section{Summary}\label{conc}

Let us summarize the main results achieved in this work.

We have presented two heuristic methods to get tight two-party two-outcome Bell inequalities. Using
these methods we have extended the list of known 4422 type inequalities by 129 members. The list is
very probably still not full.

For 241 inequalities we used numerical optimization to determine measurement operators and states
giving their maximum quantum violation achievable with six-dimensional complex component Hilbert
spaces, allowing three-dimensional projectors as measurement operators, and with eight-dimensional
real component Hilbert spaces, allowing four-dimensional projectors as measurement operators. An
$n$-dimensional calculation with $(n/2)$-dimensional projectors covers all $m$-dimensional ($m\le n$)
cases with projectors having any dimensionality between $\max(0,m-n/2)$ and $\min(m,n/2)$. The 241
examples we have considered include all tight bipartite Bell inequalities with up to five two-outcome
measurement settings known to us (excluding CHSH, about which we could tell nothing new). These
results represent lower limits for the quantum bounds for the maximum violation. At the same time, the
method proposed in Refs.~\cite{NPA07,NPA08} makes it possible to determine a series of upper bounds by
carrying out the calculation through a hierarchy of levels. The series converges to the exact value,
often reaching it at a finite level. For the majority of cases the upper bound has become equal to the
lower one at a level which is quite easy to perform for inequalities with no more than five settings
per party. There remained only 20 cases for which we still can not tell for sure the maximum value of
their quantum violation. The lower limit is probably the exact value for quite a few of them, but
unfortunately we could not do the calculation for the upper limit at a high enough level to verify it.
It is surprising that for the smallest case we considered, with only three measurement settings per
party, there is still a significant gap between the upper and lower bounds, while the former seems to
come only marginally lower by increasing the level, and for the latter we were unable to find any
better solution than one can get with real qubits, despite the high dimensionality of the component
spaces we considered and the many attempts we have done. We have only found a few inequalities whose
maximum violation was achieved with the maximally entangled state, in all those cases a pair of qubits
were sufficient.

We have also calculated the minimum detection efficiency to verify quantum violation for each
inequality with a maximally entangled pair of particles, both for symmetric and asymmetric
arrangements. By allowing degenerate measurements (ones with a certain outcome), one could lower these
threshold efficiencies for most inequalities, but this only means that the (not necessarily tight)
inequality with less measurement settings one can get by skipping the degenerate ones is at least as
good at verifying quantum violation. As far as the actual numbers are concerned, for the symmetric
case none of the inequalities does better than the best one already known, while for the asymmetric
case we have found three inequalities with five measurement settings per party, for which slightly
less efficient detectors are sufficient than for the best one known so far.

\acknowledgments Many thanks to Stefano Pironio for pointers to
the literature. T.V. has been supported by a J\'anos Bolyai Grant
of the Hungarian Academy of Sciences.

\end{document}